\documentclass[prd,twocolumn,nofootinbib,longbibliography,superscriptaddress]{revtex4-1}

\usepackage{graphicx}
\usepackage{dcolumn}
\usepackage{bm}
\usepackage{amsmath}
\usepackage{amsfonts}
\usepackage{dsfont}
\usepackage{upgreek}
\usepackage{braket}
\usepackage{float}
\usepackage{hyperref}
\usepackage[dvipsnames]{xcolor}

\begin{document}

\title{Dark energy effects in the Schr\"odinger-Newton approach}

\author{Kelvin}
\affiliation{School of Physical and Mathematical Sciences, Nanyang Technological University, Singapore 637371, Singapore}

\author{Kelvin Onggadinata}
\affiliation{School of Physical and Mathematical Sciences, Nanyang Technological University, Singapore 637371, Singapore}

\author{Matthew J. Lake}
\affiliation{School of Physics, Sun Yat-Sen University, Guangzhou 510275, People's Republic of China}
\affiliation{School of Physical and Mathematical Sciences, Nanyang Technological University, Singapore 637371, Singapore}

\author{Tomasz Paterek}
\affiliation{School of Physical and Mathematical Sciences, Nanyang Technological University, Singapore 637371, Singapore}
\affiliation{MajuLab, International Joint Research Unit UMI 3654, CNRS, Universit{\' e} C\^ote d'Azur, Sorbonne Universit{\' e}, National University of Singapore, Nanyang Technological University, Singapore}
\affiliation{Institute of Theoretical Physics and Astrophysics,
Faculty of Mathematics, Physics and Informatics,
University of Gda\'nsk, 80-308 Gda\'nsk, Poland}

\date{\today}

\begin{abstract}
The Schr\"odinger-Newton equation is a proposed model to explain the localization of macroscopic particles by suppressing quantum dispersion with the particle's own gravitational attraction.
On cosmic scales, however, dark energy also acts repulsively, as witnessed by the accelerating rate of universal expansion.
Here, we introduce the effects of dark energy in the form of a cosmological constant $\Lambda$, that drives the late-time acceleration of the Universe, into the Schr\"odinger-Newton approach.
We then ask in which regime dark energy dominates both canonical quantum diffusion and gravitational self-attraction.
It turns out that this happens for sufficiently delocalized objects with an arbitrary mass and that there exists a minimal delocalization width of about $67$ m.
While extremely macroscopic from a quantum perspective, the value is in principle accessible to laboratories on Earth.
Hence, we analyze, numerically, how the dynamics of an initially spherical Gaussian wave packet is modified in the presence of $\Lambda > 0$. 
A notable feature is the gravitational collapse of part of the wave packet, in the core region close to the center of mass, accompanied by the accelerated expansion of the more distant shell surrounding it.
The order of magnitude of the distance separating collapse from expansion matches analytical estimates of the classical turnaround radius for a spherically symmetric body in the presence of dark energy.
However, the time required to observe these modifications is astronomical.
They can potentially be measured only in physical systems simulating a high effective cosmological constant, or, possibly, via their effects on the inflationary universe.
\end{abstract}

\maketitle

\section{\label{sec:level1}Introduction}

Quantum theory places no restrictions on the extent to which particles can be superposed in space, yet macroscopic objects are never observed delocalized.
A number of solutions to this puzzle have been suggested both within the canonical quantum formalism, e.g., decoherence \cite{Zurek-PT,dec-book} and the theory of coarse-grained measurements \cite{Peres-book,Poulin1995,KB2007,KB2008,Navascues,Macro-corrs}, as well as outside it, e.g., the so-called collapse models \cite{bassi2013models}.
The latter class includes theories with no free parameters in which gravitational self-interaction counters the spread of the quantum wave function, effectively ``collapsing'' quantum superpositions~\cite{Karolyhazy1966, diosi1984gravitation, penrose1996gravity, bassi2013models, moller1962theories, rosenfeld1963quantization}. 
One such way of coupling gravity to the quantum description is the semiclassical approach proposed by M{\o}ller \cite{moller1962theories} and Rosenfeld \cite{rosenfeld1963quantization} that results in the Schr\"odinger-Newton (SN) equation in the static weak-field limit \cite{diosi1984gravitation}.
In this model, even a single massive quantum particle interacts with the mass distribution that it generates.
The modified Schr\"odinger equation involves the Newtonian gravitational potential with a mass distribution $\rho_{m} = m|\psi|^2$, where $m$ is the mass of the particle and 
$|\psi|^2$ is the probability density of finding it in a particular spatial location, according to the standard Born rule.

Here, we extend the SN approach to include the effects of dark energy in the form of a positive cosmological constant, $\Lambda >0$.
In 1998, redshift observations of type Ia supernovae found that our Universe is expanding at an accelerating rate~\cite{riess1998observational, perlmutter1999measurements} and a wealth of cosmological data, including analyses of the large-scale structure \cite{Betoule:2014frx}, baryon acoustic oscillations \cite{Heavens:2014rja}, and the cosmic microwave background (CMB) radiation \cite{aghanim2018planck}, now support this view. 
Dark energy, an unknown form of energy permeating the whole of space, is suspected to account for this expansion. 
Although many models of dark energy exist in the literature~\cite{dark-review}, we focus here on the simplest one, known as the Lambda cold dark matter ($\Lambda$CDM) ``concordance'' model~\cite{Ostriker:1995rn}.
This includes an additional cosmological constant term ($\propto \Lambda g_{\mu\nu}$) in the gravitational field equations derived from the usual Einstein-Hilbert action \cite{wald1984general}. 
In the static weak-field limit, this contributes an additional (repulsive) term to the (attractive) Newtonian potential of a compact spherically symmetric body. 
The new term is proportional to $\Lambda$ and to the square of the radial distance, i.e., $\Phi_{\Lambda} \propto \Lambda r^2$~\cite{Hobson:2006se}.

A particle described by the resulting Schr\"odinger-Newton-de Sitter (SNdS) equation, which we introduce below, qualitatively experiences three tendencies in its dynamics: 
its wave packet tends to spread as a result of both canonical quantum dispersion and dark energy, whereas Newtonian gravity tends to localize the object (The nonrelativistic limit of the evolution of a massive scalar field in dS space was also considered in \cite{Fodor:2010hg} but only small oscillatory solutions were studied. These qualitatively resemble the solutions obtained in the oscillatory regime of the SNdS system, considered here (see Table \ref{tab: SN results}). A similar analysis of small-amplitude oscillations in the nonrelativistic limit of the Einstein-massive-scalar system in AdS space, i.e., with $\Lambda < 0$, was performed in \cite{Bizon:2018frv}. In the present analysis, we go beyond the study of small oscillations to consider the full evolution of the SNdS system under various regimes).
We conduct an analysis similar to Carlip's~\cite{carlip2008quantum}, which compares the effects of these localizing and delocalizing tendencies, and which allows us to estimate the experimental parameters for which the dark energy contribution dominates.
This results in a clear requirement, namely that the macroscopic spread of the wave function must be no less than $67$ m, regardless of the particle's mass.
We stress that the analysis puts no limits on the mass of the particle; i.e., the quantum dynamics of any particle wave function could (in principle) manifest observable effects due to dark energy repulsion.
In particular, for an object of mass of the order of $10^{-20}$ kg, dark energy dominates already for the initial width of $67$ m.

Of course, quantum experiments at such macroscopic scales are extremely difficult, but there is no fundamental reason against their execution in a very isolated laboratory.
This hints at an in principle possibility to observe dark energy effects within a terrestrial setup.
Hence, we numerically compute the evolution of an initially spherical Gaussian wave function (with an initial width of $75$ m), according to the SNdS equation, 
and systematically study deviations of the particle dynamics from both SN and canonical quantum evolution.
The results are sensitive to the mass of the particle and demonstrate both a faster-than-quantum expansion of the wave packet 
as well as a gravitational collapse of the wave packet core, together with an accelerated expansion of the outer shell, for particles beyond a threshold mass of order $2 \times 10^{-21} \, {\rm kg}$.
Estimates of the separation distance between the collapsing and expanding parts are given in terms of the classical turnaround radius, which are in harmony with our numerical results.

Unfortunately, all these effects become clearly distinguishable from the canonical and SN dynamics only after the particle has evolved for astronomically long durations.
We therefore hope that our analysis will stimulate discussions on shortening the required time and note that these effects may (again, in principle) be simulated in analogue gravitational systems;
see e.g.,~\cite{FF2003,Weinfurtner2005,FLS2012,JV2012,BLM2014,DLT2016} for the analogues of the expanding universe.
Since a bigger effective cosmological constant gives rise to more noticeable differences at earlier epochs, our analysis may also be relevant to studies of quantum perturbations generated by cosmological inflation \cite{Liddle:2000cg}.

\section{The Schr\"odinger-Newton-de Sitter Equation}

We begin by extending the SN equation to include the cosmological constant.
The treatment assumes a single particle of mass $m$ in the presence of dark energy, in the form of a constant energy density sourced by $\Lambda > 0$.
Following M{\o}ller \cite{moller1962theories} and Rosenfeld \cite{rosenfeld1963quantization}, our starting point is the semiclassical formulation of the Einstein field equations,
\begin{align}\label{eq: SNLambda EFE}
    G_{\mu \nu} + \Lambda g_{\mu \nu} = \frac{8\pi \mathcal{G}}{c^{4}} \bra{\psi} \hat{T}_{\mu \nu} \ket{\psi} \, . 
\end{align}
Here, $G_{\mu \nu} = R_{\mu\nu} + (1/2)R \, g_{\mu\nu}$ is the Einstein tensor, $R_{\mu\nu}$ is the Ricci tensor, $R = R^{\mu}{}_{\mu}$ is the scalar curvature, and $g_{\mu \nu}$ is the metric tensor. 
We denote Newton's constant by $\mathcal{G}$, and $\hat{T}_{\mu \nu}$ is the energy-momentum tensor operator,
whose expectation value in the state $| \psi \rangle$, describing the particle, enters on the right-hand side of Eq. (\ref{eq: SNLambda EFE}).
This is interpreted as treating the gravitational field classically while matter is described quantum mechanically. 
Taking the static weak-field limit, we execute steps analogous to those leading to the Newtonian limit (which is recovered by setting $\Lambda = 0$) found in textbooks on general relativity~\cite{Hartle-book}. 
This reduces Eq.~(\ref{eq: SNLambda EFE}) to the following equation for the effective potential $\Phi (\vec{r},t)$:
\begin{align}
     \nabla^{2} \Phi (\vec{r},t) = 4\pi \mathcal{G}m |\psi(\vec{r},t)|^{2} - \Lambda c^{2} \, .
     \label{EQ_POISSON}
\end{align}
That is, $m |\psi(\vec{r},t)|^{2}$ plays the role of the mass distribution in the usual Poisson's equation for the Newtonian potential and the second term arises due to dark energy. 
The latter is independent of the sourcing mass, in agreement with its origin as a property of space \cite{Hobson:2006se}.
Equation~(\ref{EQ_POISSON}) admits the following solution:
\begin{align}\label{eq: SNLambda potential}
    \Phi(\vec{r},t) = -\mathcal{G}m \int \frac{|\psi(\vec{r} \, ',t)|^{2}}{|\vec{r} - \vec{r} \, '|}d^{3}\vec{r} \,' + \frac{\Lambda c^{2}}{4 \pi} \int \frac{1}{|\vec{r} - \vec{r} \, '|}d^{3}\vec{r} \, ' .
\end{align}
Finally, we define the SNdS equation as the Schr\"odinger equation for our quantum particle of mass $m$, with the effective potential introduced above, i.e.,
\begin{align}
\label{EQ_SNDS}
i \hbar \frac{\partial}{\partial t} \psi(\vec{r},t) =  \left( -\frac{\hbar^{2}}{2m} \nabla^{2} + m \Phi(\vec{r},t) \right) \psi(\vec{r},t) \, , 
\end{align}
where $\Phi(\vec{r},t)$ is given by Eq. (\ref{eq: SNLambda potential}). 
Clearly, by setting $\Lambda = 0$, we recover the standard SN equation.
The particle is said to self-interact as, in this model, it propagates within a potential generated by itself.
This leads to nonlinear evolution of the quantum state, which is explicitly visible after plugging (\ref{eq: SNLambda potential}) into (\ref{EQ_SNDS}).
Hence, we must deal with a complicated integro-differential equation of motion, with little hope for analytical exploration, though see Ref.~\cite{Mozor1998} for attempts in this direction.
Therefore, in order to compute the dynamics of an object evolving under the SNdS equation, we closely follow the numerical method presented in the thesis by Salzman~\cite{salzman2005investigation}.

For simplicity, the initial state is chosen to be a three-dimensional spherically symmetric Gaussian wave function,
\begin{align}\label{eq: initial gaussian wavefunction}
    \psi(r,0) = \left( \frac{\alpha}{\pi} \right)^{3/4} e^{-\alpha r^{2}/2},
\end{align}
where the initial width $\sigma$ is given by $\sigma = \alpha^{-1/2}$ and $r$ is the radial distance.
Under spherical symmetry, the effective potential (\ref{eq: SNLambda potential}) also depends only on $r$ and Eq. (\ref{EQ_SNDS}), which must be solved numerically, contains only one variable.
However, the effects induced by the cosmological constant are typically hard to observe due to its extremely small value.
We recall that the numerical value of the cosmological constant, inferred from observations of type Ia supernovae \cite{riess1998observational, perlmutter1999measurements}, large-scale structure \cite{Betoule:2014frx}, and the CMB \cite{aghanim2018planck}, is $\Lambda = 1.089 \times 10^{-52}$ m$^{-2}$.
It would therefore be useful to have an estimate of the size of the initial width and the mass of the particle for which $\Lambda$ can produce noticeable changes in $\psi(r,t)$.
We now provide such an estimate by following an argument analogous to that introduced by Carlip~\cite{carlip2008quantum}, but with $\Lambda >0$.

\section{Estimation of the minimal width}

\begin{figure*}[!t]
\includegraphics[width=1.0\linewidth]{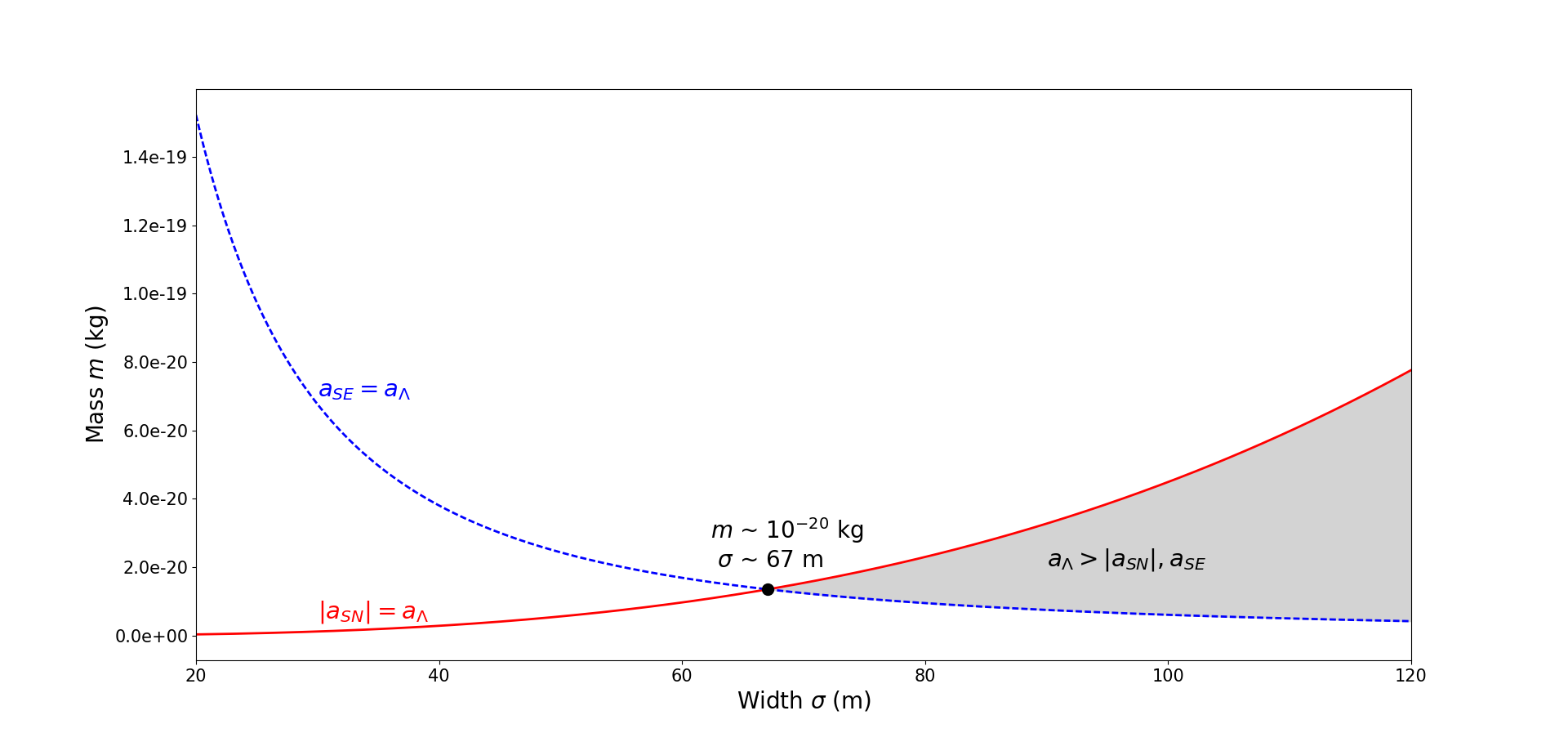}
\caption{The minimal width of the initial wave function required for dark energy to dominate quantum diffusion and gravitational self-attraction.
The blue dashed line gives the relation between the mass of the particle and the initial width for which the (repulsive) accelerations due to the canonical quantum spread and due to dark energy are of the same order of magnitude, i.e., $a_\Lambda = a_{\mathrm{SE}}$.
Dark energy dominates above this curve.
Similarly, the acceleration due to dark energy dominates gravitational acceleration below the red solid line, which is obtained by setting $a_\Lambda = |a_{\mathrm{SN}}|$.
The shaded region shows where $a_\Lambda$ is higher than the absolute values of the other accelerations.
This only happens for an initial width above $\sim 67$ m.
}
\label{fig: accelerations plot}
\end{figure*}

In canonical quantum theory, the initial spherical Gaussian wave function spreads outwards under free evolution according to the Schr\"odinger equation.
Let us denote by $r_p$ the position of the spherical shell at which the radial probability density has its maximum, i.e., the radius at which the particle is most likely to be found.
The canonical spread is characterized by the ``accelerating'' peak of the radial probability density, $\ddot r_p$,
\begin{equation}
a_{\mathrm{SE}} \sim \frac{\hbar^2}{m^2 r_p^3} \, .
\label{EQ_A_SE}
\end{equation}

In the SN approach, there is a competing tendency due to gravitational self-interaction which can be estimated as gravitational acceleration towards the center ($r=0$) at the point $r_p$,
\begin{equation}
a_{\mathrm{SN}} \sim -\frac{\mathcal{G}m}{r_{p}^{2}} \, .
\label{EQ_A_SN}
\end{equation}
Here, we assume that essentially all the mass is within a radius of order $r_p$ and note that slightly better accuracy is obtained from Gauss's law by taking into account the actual fraction of the mass within the range $0 < r \leq r_p$ (see Appendix~\ref{APP_GAUSS}).

Finally, from the dark energy part of the effective potential, one obtains the outward acceleration given by
\begin{equation}
a_{\Lambda} \sim \frac{1}{3} \Lambda c^{2} r_{p}\, .
\label{EQ_A_L}
\end{equation}
Combining these three estimates, and keeping in mind that, at $t = 0$, the peak of the radial probability density, $r_p$, is equal to the width $\sigma$ of the spherical Gaussian,
we can now estimate the regime in which the acceleration due to dark energy dominates the accelerations due to both Newtonian gravity and canonical quantum diffusion.
Figure~\ref{fig: accelerations plot} shows the mass-width relations for pairwise accelerations to be of the same order of magnitude.
The main finding is that there exists a critical initial Gaussian width such that the acceleration induced by the cosmological constant dominates for a certain mass range. 
This critical width is approximately $67$ m. 

Therefore, one expects a Gaussian wave function with an initial width greater than the critical value to evolve identically to that of a free particle for small masses, since the acceleration due to quantum dispersion dominates in this region. 
As the mass is increased, the acceleration due to dark energy starts to dominate canonical quantum diffusion and, finally, for even larger masses, gravitational acceleration dominates and the evolution reduces to the SN case.
With this in mind, we now study the SNdS evolution of a wave function with an initial width of $75$ m, corresponding to $\alpha = 1.78 \times 10^{-4} \, {\rm m^{-2}}$.

\section{Dark energy effects}

We have validated our numerical approach by performing simulations of the SN equation in the regimes previously studied in the literature.
All the results are in good agreement with Ref.~\cite{giulini2011gravitationally}.
For completeness, these are summarized in Appendix~\ref{APP_SN}, where we also discuss the oscillating solutions in more detail and give an intuitive physical explanation for their appearance in the SN and SNdS models.

In short, the wave function of a particle obeying the SN equation undergoes one of three possible, qualitatively different, evolutions. 
It either spreads more slowly than the wave function of a canonical quantum particle of the same mass, or its peak radial probability oscillates about an equilibrium distance, or, finally, it collapses gravitationally towards the center of mass. 
We note that, even in the SN model (with $\Lambda = 0$) the time scale of gravitational collapse, for particles with masses of order $10^{-20}$ kg and initial width of $75$ m,
is of the order of $10^{17}$ seconds, i.e., comparable to the de Sitter time scale, $t_{\rm dS} = c^{-1}\sqrt{3/\Lambda} \sim 10^{17} \, {\rm s}$, which in turn is comparable to the present age of the Universe \cite{Hobson:2006se}.

\begin{figure}[!b]
\centering
\includegraphics[scale=0.31]{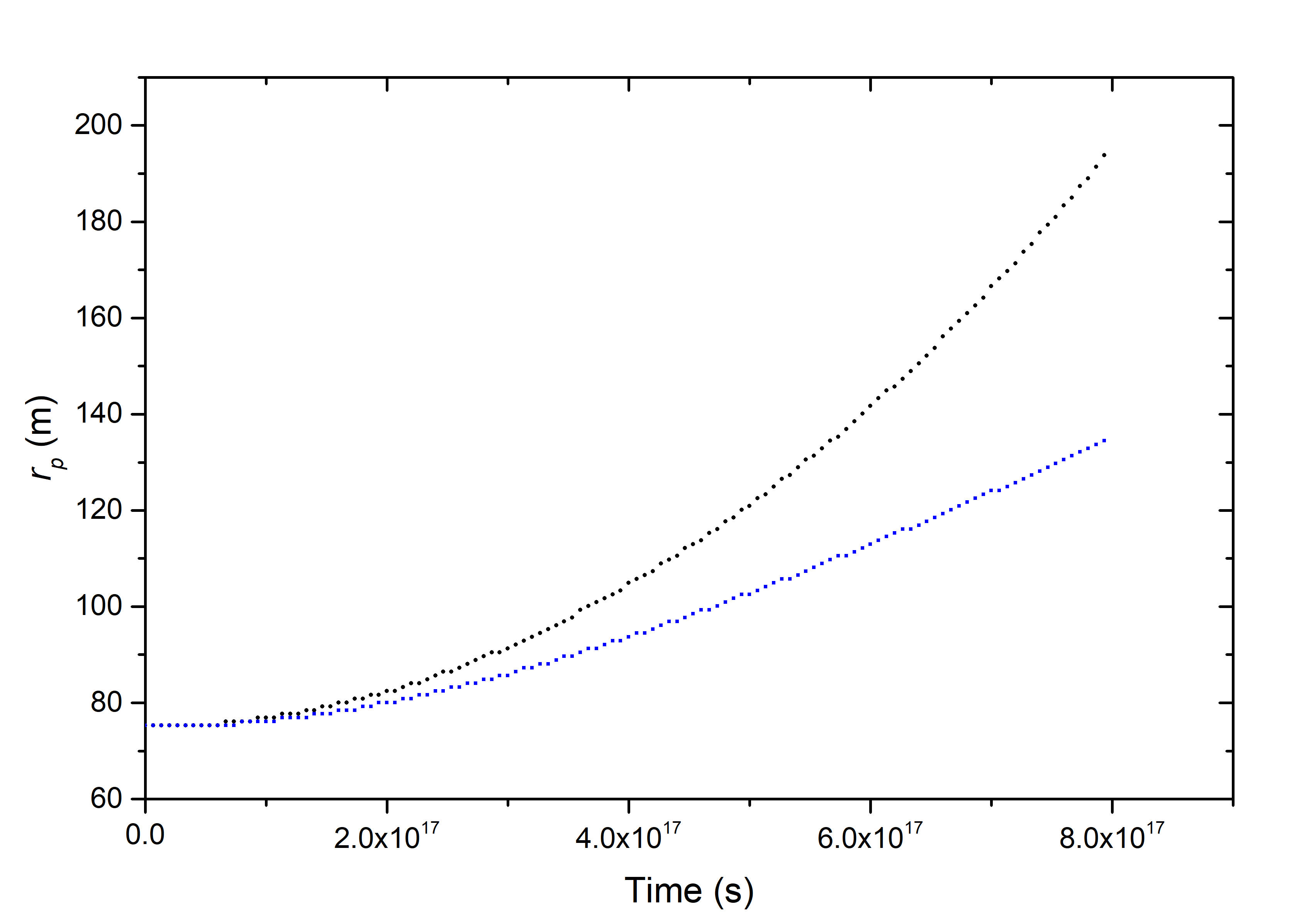}
\caption{Excessive expansion of the wave function in the presence of dark energy.
We numerically simulated the SNdS evolution of a spherically symmetric Gaussian wave packet with an initial width $\sigma = 75$ m, for a particle of mass $m = 1 \times 10^{-20}$ kg.
The black dots give the position of the peak radial probability density for a particle evolving under the SNdS equation.
The blue dots follow the analytical solution for the peak probability density of a free particle, with the same parameters, evolving under the Schr\"odinger equation.
Note the astronomical time scale.}
\label{fig: SNLambda m=1E-20}
\end{figure}

In the presence of a positive cosmological constant, the SNdS equation gives rise to two effects that are distinct from both canonical quantum theory and evolution under the SN equation.
Table~\ref{tab: SNLambda results} systematically summarizes the obtained numerical results.
The first effect is present for masses in the range $\sim 2 \times 10^{-21}$ kg to $\sim 3 \times 10^{-20}$ kg and shows that they spread faster than the corresponding free quantum particle,
i.e., the peak of the radial probability density moves away from the origin faster than the canonical prediction. 
Since the spread in SN dynamics is never as fast as in the canonical theory, this effect is due to dark energy.
In Fig.~\ref{fig: SNLambda m=1E-20}, we present results of numerics that clearly show the excessive expansion of the wave packet in the presence of the cosmological constant.
Note that the time scale for the onset of this characteristic effect is also comparable to the de Sitter time and hence, for $\Lambda \sim 10^{-52} \, {\rm m^{-2}}$, to the present age of the Universe $\sim 10^{17} \, {\rm s}$.

\setlength{\tabcolsep}{6pt}
\renewcommand{\arraystretch}{1.7}
\begin{table}[!t]
\centering
\caption{Features of dynamical evolution under the SNdS equation with an initially spherically symmetric Gaussian wave function of width $\sigma = 75$ m ($\alpha = 1.78\times 10^{-4}$ m$^{-2}$).
The comparison is made to a free particle in canonical quantum mechanics.
}
\label{tab: SNLambda results}
    \begin{tabular}{p{4cm}p{3.8cm}}
        \hline
        Mass                         & Behavior                                \\ \hline \hline
        Below $1 \times 10^{-21}$ kg & Identical to the free particle \\ \hline
        $2\times 10^{-21}$ kg to $3\times 10^{-20}$ kg                           & Spreads faster than the free particle    \\ \hline
        $4 \times 10^{-20}$ kg to $5 \times 10^{-20}$ kg                           & Inner core of the wave function spreads slower than the free particle while the other shell spreads faster                          \\ \hline
        $6\times 10^{-20}$ kg to $1 \times 10^{-19}$ kg                         & Inner core of the wave function collapses under self-gravity while the outer shell spreads faster than in canonical quantum mechanics  \\ \hline
	 $ \sim 2 \times 10^{-19}$ kg & Chaotic \\ \hline
        Above $3 \times 10^{-19}$ kg                           & Stationary                               \\
        \hline
    \end{tabular}
\end{table}

For masses above $4 \times 10^{-20}$ kg, the gravitational interaction starts to play a significant role in the evolution of the wave function. 
Since the Newtonian potential is inversely proportional to the radial distance, whereas the dark energy potential is proportional to square of the radial distance,
canonical gravitational attraction dominates for small radii while the repulsive effects of dark energy become significant at large radii. 
Therefore, given a Gaussian wave packet with a sufficiently large initial width, one expects a core region of the wave packet, within some critical radius close to the center of mass, to collapse, while the surrounding outer shell, which lies outside the critical radius, continues to expand away from the center.
This is exactly what we observe in the numerics up to masses of order $1 \times 10^{-19}$ kg.

\begin{figure}[!t]
\centering
\includegraphics[scale=0.42]{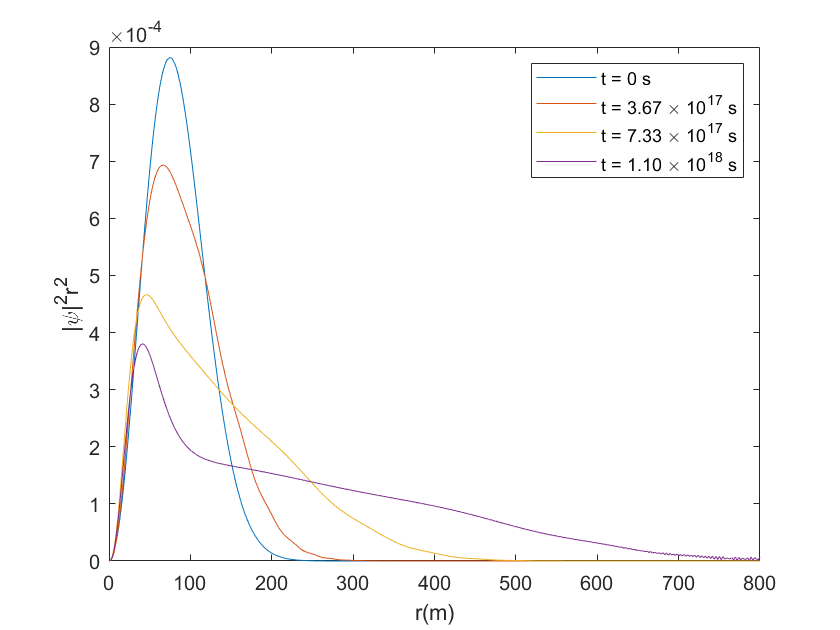}
\caption{Combined effect of gravity and dark energy.
The plot presents radial wave functions at four time instances for a particle with an initial spread $\sigma = 75$ m and mass $m = 5 \times 10^{-20}$ kg.
Part of the wave function near the origin gravitationally collapses whereas the part further away expands due to dark energy.
The boundary is at a distance of about $100$ m and matches the classical turnaround radius, as explained in the text.
Note the astronomical time scale.
}
\label{fig: SNLambda m=5E-20}
\end{figure}

Furthermore, the inner core exhibits the same (qualitative) dynamics as observed in the SN model, whereas the parts of the wave packet lying at relatively large radial distances tend to spread faster than in the canonical free particle case.
An example of such an evolution is given in Fig.~\ref{fig: SNLambda m=5E-20}.
We note that, for masses above $6 \times 10^{-20}$ kg, the peak of the collapsing radial probability density oscillates about an equilibrium position, in full analogy to the SN case. 
An in depth analysis of this result is presented in Appendix~\ref{APP_SN}.
In general, we find that the larger the mass, the larger the fraction of the wave function that collapses.
One would therefore expect that, for a sufficiently large mass, the fraction of the wave packet that spreads away would vanish, so that the whole system undergoes gravitational collapse, 
i.e., that evolution under the SNdS equation reduces to SN evolution for sufficiently large $m$. 
However, unfortunately, this regime cannot be reached with our present numerical procedures.

We also would like to comment that the features listed in the last two rows of Table~\ref{tab: SNLambda results} are most likely artifacts of the numerics.
By ``chaotic'', we mean that it is not possible to describe the dynamical evolution in simple terms 
whereas in the ``stationary'' regime, we observe no changes of the wave function within the simulation time.
Similar effects (also considered to be numerical artifacts) were observed by Salzman~\cite{salzman2005investigation}.

Finally, we explain the origin of the critical distance in Fig.~\ref{fig: SNLambda m=5E-20} which demarcates the boundary between collapse and expansion.
In the presence of dark energy, there exists a maximum radius around a spherical mass distribution, outside of which a probe mass is always repelled. 
This is the region in which the repulsive effects of dark energy dominate the Newtonian gravitational attraction.
For classical systems, this maximum stable radius is known as the turnaround radius \cite{bhattacharya2017maximum}. 
In the case of the Schwarzschild-de Sitter spacetime, which represents the gravitational field around a quantum ``particle" in our model, it is given by
\begin{align}
    r_{\leftrightarrow} = \left( \frac{3\mathcal{G}M}{\Lambda c^{2}} \right)^{1/3} \, .
\end{align}
This expression can also be obtained, in the weak-field limit, by simply equating $|a_{\mathrm{SN}}|$ and $a_{\Lambda}$, given by Eqs. (\ref{EQ_A_SN}) and (\ref{EQ_A_L}), respectively.
Similarly, one can introduce a quantum mechanical turnaround radius from the requirement that the outward acceleration, being the sum $a_{\mathrm{SE}} + a_{\Lambda}$, balances the inward acceleration $|a_{\mathrm{SN}}|$.
However, since $a_{\mathrm{SE}}$ decays with the radial distance faster than the other accelerations,
the classical turnaround radius is expected to be a good estimate (or small overestimate) of the distance above which the dark energy dominates, also for quantum wave functions.
Indeed, for a mass of $m = 5 \times 10^{-20}$ kg, the classical turnaround radius is $100$ m, in good agreement with Fig.~\ref{fig: SNLambda m=5E-20}.

\section{Conclusions}

We have incorporated dark energy in the form of a cosmological constant in the Schr\"odinger-Newton approach.
In our model, the dynamics of a quantum particle is determined by a nonlinear Schr\"odinger-like equation, with an effective potential including both Newtonian self-gravity and dark energy contributions. 
We dubbed this the Schr\"odinger-Newton-de Sitter (SNdS) equation. 
We then estimated the particle mass, as well as the initial width of a spherically symmetric Gaussian wave function, for which the dark energy contribution dominates both canonical quantum diffusion and Newtonian self-interaction.
Surprisingly, we found that, while there are no fundamental restrictions on the mass, the initial width must be larger than a critical value of approximately $67$ m.
If this condition is satisfied, two distinctive phenomena witnessing dark energy emerge in the resulting dynamics.
The first is the faster-than-quantum spread of the wave function, and the second combines gravitational collapse of the inner core of the wave packet, close to the center of mass, with the accelerated spread of the outer shell, lying outside a given critical radius.
In addition, we showed that the boundary between the collapsing and expanding parts is well estimated by the classical turnaround radius.

Unfortunately, both of these effects take an astronomically long time to accumulate observable differences between SNdS and canonical quantum or SN-like evolution.  
According to our numerics, the required time is of the order of the age of the Universe, which is comparable to the de Sitter time scale, $t_{\rm dS} = c^{-1}\sqrt{3/\Lambda} \sim 10^{17} \, {\rm s}$. 
This can also be understood analytically since, in the asymptotic de Sitter phase corresponding to the present rate of universal expansion, the Friedmann equation determining the evolution of the scale factor of the Universe reduces to $\ddot{a}(t) \simeq c^{2}(\Lambda/3)a(t)$ \cite{Islam:1992nt}. 
This is easily solved to obtain $a(t) \propto \exp\left(t/t_{\rm dS}\right)$. 
Similarly, Eq. (\ref{EQ_A_L}) may be rewritten as $\ddot{r}_p(t) \simeq c^{2}(\Lambda/3)r_p(t)$, yielding $r_p(t) \propto \exp\left(t/t_{\rm dS}\right)$ for the late-time evolution of $r_p(t)$, in good agreement with the results presented in Fig 2. 

Nevertheless, it is possible that these effects could be simulated in analogue gravity systems with larger effective cosmological constants; see e.g.,~\cite{FF2003,Weinfurtner2005,FLS2012,JV2012,BLM2014,DLT2016}.
There may also exist precision experiments which considerably reduce the time required, 
though it seems challenging to design a scheme capable of revealing the presence of $\Lambda$ in a terrestrial laboratory within the time-frame of, say, one human generation.

\begin{acknowledgements}

We thank Sri Devi Wijaya for discussions.
This work was supported by Singapore Ministry of Education Academic Research Fund Tier 1 Project No. RG106/17
and Polish National Agency for Academic Exchange NAWA Project No. PPN/PPO/2018/1/00007/U/00001.
Both Kelvins acknowledge support from Nanyang Technological University under the Undergraduate Research Experience on Campus (URECA) program. 
ML thanks Nanyang Technological University for hospitality during the preparation of the manuscript.

\end{acknowledgements}

\appendix

\section{Correction to the gravitational acceleration in the SN approach}
\label{APP_GAUSS}

The expression for the Newtonian gravitational acceleration, Eq. (\ref{EQ_A_SN}) in the main text, assumes that all the mass sources the gravitational force.
But, according to Gauss's law, only the mass within the radius $r_p$ is relevant for the spherically symmetric mass distributions considered here.
This effective mass is therefore given by $\zeta m$, where
\begin{align}
    \zeta = 4 \pi \int_{0}^{r_{p}} |\psi(r,0)|^2 r^2 dr \, . 
\end{align}
In Appendix~\ref{APP_SN}, we use this effective mass to improve our estimate of the equilibrium distance in the SN dynamics.

\section{Validation of our numerics: reproducing SN evolution with $\Lambda = 0$}
\label{APP_SN}

In order to validate our numerics, we have simulated the SN equation for the initial spherically symmetric Gaussian wave packet with width $\sigma = 0.5$ $\upmu$m ($\alpha = 4 \times 10^{12}$ m$^{-2}$).
The results are in good agreement with those obtained in Ref.~\cite{giulini2011gravitationally} (see also \cite{SC2006}).
A summary of the observed features for different masses is given in Table \ref{tab: SN results}.

\setlength{\tabcolsep}{6pt}
\renewcommand{\arraystretch}{1.7}
\begin{table}[!htb]
\centering
\caption{Observed dynamics of a particle initially in a spherically symmetric Gaussian state of a width $\sigma = 0.5$ $\upmu$m, for different masses, according to the SN equation.
The comparison is made to the case of a free particle in canonical quantum theory.
The oscillatory behavior refers to oscillations of the peak of the radial probability density; see Fig.~\ref{fig: SN m=2E-17}.}
\label{tab: SN results}
    \begin{tabular}{p{4cm}p{3.6cm}}
        \hline
        Mass                         & Behavior                                \\ \hline \hline
        Below $3 \times 10^{-18}$ kg & Identical to the free particle \\ \hline
        $3\times 10^{-18}$ kg to $1\times 10^{-17}$ kg      & Spreads slower than the free particle    \\ \hline
        $\sim 2 \times 10^{-17}$ kg                           & Oscillates \\ \hline
        $3\times 10^{-17}$ kg to $9 \times 10^{-17}$ kg     & Collapses towards the center of mass    \\ \hline
        $1 \times 10^{-16}$ kg to $4 \times 10^{-16}$ kg    & Chaotic                                  \\ \hline
        Above $5 \times 10^{-16}$ kg                        & Stationary                               \\ \hline
    \end{tabular}
\end{table}

\begin{figure}[!htb]
\centering
\includegraphics[scale=0.42]{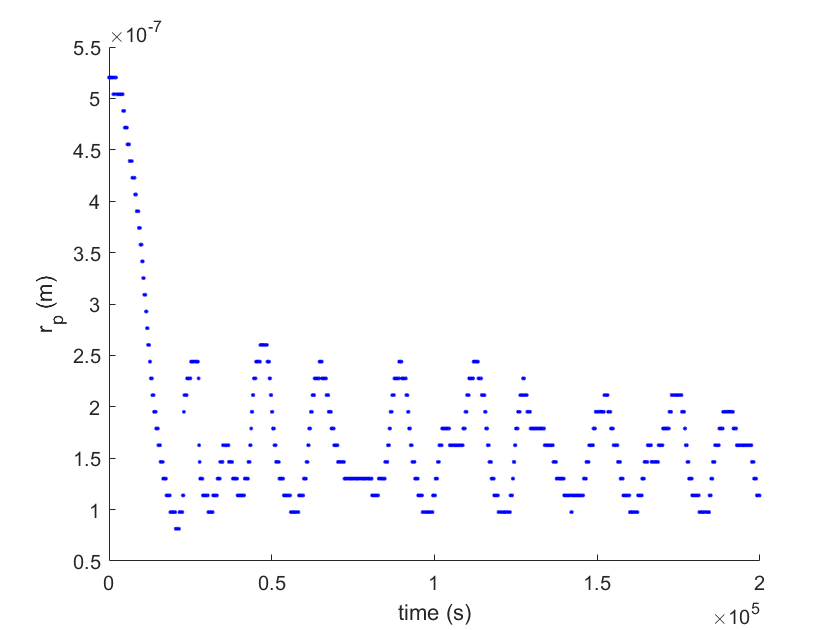}
\caption{The evolution of the peak of the radial probability density for a particle of mass $2 \times 10^{-17}$ kg evolving under the SN equation.}
\label{fig: SN m=2E-17}
\end{figure}

Let us briefly discuss the ``oscillating" behavior of the wave function, observed for $m \sim 2 \times 10^{-17}$ kg; see Fig.~\ref{fig: SN m=2E-17}.
Qualitatively, one expects this as a result of the unstable equilibrium between the accelerations $a_{\rm SE}$ (\ref{EQ_A_SE}) and $a_{\rm SN}$ (\ref{EQ_A_SN}).
Indeed, the outward acceleration $a_{\rm SE}$ dominates for small distances whereas the inward acceleration $a_{\rm SN}$ dominates at large distances.
The order of magnitude of the equilibrium distance is given by $r_{\text{eq}} \sim \hbar^{2}/\mathcal{G} m^{3}$.
This implies that, for large masses, $r_{\text{eq}}$ is extremely close to the origin and gravitational acceleration dominates nearly everywhere in space. 
On the other hand, for small masses, $r_{\text{eq}}$ is extremely far away from the origin and the outward acceleration dominates nearly everywhere in space causing the wave packet to spread indefinitely.
For a mass of $m = 2 \times 10^{-17}$ kg, this estimate gives $r_{\text{eq}} = 2 \times 10^{-8}$ m, which is 1 order of magnitude away from the numerical finding.
However, not all of the mass contributes to the acceleration at $r_p$ and, following Appendix~\ref{APP_GAUSS}, we obtain an effective mass of $8.6 \times 10^{-18}$ kg, together with the corresponding radius of equilibrium, $r_{\text{eq}} = 2.62 \times 10^{-7}$ m.

\bibliography{references}

\end{document}